\documentstyle[prb, aps, epsf]{revtex}
\begin{document}
\draft

\title{Relation between dispersion lines and conductance
 of telescoped armchair double-wall nanotubes analyzed using perturbation formulas and first-principles calculations}

\author{Ryo Tamura}
\address{Faculty of Engineering, Shizuoka University, 3-5-1 Johoku, 
  Hamamatsu 432-8561, Japan}
\twocolumn

\maketitle

\begin{abstract}
 The Landauer's formula conductance of 
the telescoped armchair nanotubes
 is  calculated with the Hamiltonian defined
 by first-principles calculations (SIESTA code).
 Herein, partially extracting the inner tube from the outer tube
 is called 'telescoping'.
 It shows a rapid oscillation superposed on a
 slow oscillation
 as a function of  discrete overlap length 
 $\left(L-\frac{1}{2}\right)a$ 
  with  an integer variable $L$ and  the lattice constant $a$.
  Considering the interlayer Hamiltonian
 as a perturbation,
 we obtain the approximate formula of the amplitude of the
  slow oscillation   as
 $|A|^2/(|A|^2+\varepsilon^2)$ where $A$ is the effective
 interlayer interaction and $\varepsilon$ is
 the band split without  interlayer interaction.
 The approximate formula is related to 
 the Thouless number of the dispersion lines.
\end{abstract}

\section{introduction}

 Single-wall nanotubes (SWNT) show various
 useful characteristics as nanoelectronic devices.\cite{NT-book,review}
 They can be either metallic or
 semiconducting, depending on their chirality. \cite{tubeHamada} 
 Although a semiconducting SWNT can act as 
 a transistor by itself, \cite{SWNT-transistor}
 we can increase NT device functionality 
 by assembling SWNTs.
 For example,  a junction between
 a metallic SWNT and a semiconducting
 SWNT can work as a Schottky diode.\cite{covalent-junction-MS,interlayer-junction-MS} 
 The SWNTs  can be connected by
 covalent bonds \cite{covalent-junction-MS,covalent-junction-no-MS}
and interlayer interactions.\cite{interlayer-junction-MS,interlayer-junction-no-MS} 
 Double-wall nanotubes (DWNTs), multi-wall nanotubes (MWNTs), and nanotube bundles are all examples of SWNTs that can be naturally assembled
 through interlayer interactions.
 Processing these naturally assembled SWNTs is one way
 of constructing higher-order structures for NT devices.
One prototype higher-order structure
 is the telescoped double wall nanotube (TDWNT),
 in which
 the inner tube is partially extracted
 from a DWNT, as shown in Fig.   1.
 Since the interlayer force  is relatively small,  
  the inner tube can be slid and rotated
  with respect to the outer tube by attached piezoelectric probes
 with a negligible change in intralayer bonding. \cite{experiment-telescope-mechanical,experiment-telescope-conductance} 
In this way, we can realize various interlayer configurations in the TDWNT,
 unlike the case for bulk graphite.
 When a bias voltage is applied between the attached probes, the resulting current is significantly influenced 
 by the interlayer configuration because a carrier from one probe must travel an interlayer path to  reach the other probe.
 This sensitivity 
 can be used in nanomechanical switches and nanodisplacement sensors.
 In addition, TDWNTs enable us to measure the interlayer conductance
 more directly than un-telescoped DWNTs.\cite{indirect-interlayer}

 Among the reported theoretical work on TDWNTs,\cite{theory-telescope-mechanical,theory-telescope-conductance,theory-telescope-conductance-5-5-10-10,previous-paper}, 
there are controversial results
 regarding the conductance of a TDWNT composed
 of  (10,10) and (5,5) armchair nanotubes.
 Compared to a quantum conductance $G_0=2e^2/h$, 
 the maximum conductance was found to be $2G_0$ based on tight binding (TB) model calculations,
 but only $G_0$ based on first-principles calculations.\cite{theory-telescope-conductance-5-5-10-10,previous-paper}
Landauer's formula was used in these calculations,  
 so the conductance in units of $G_0$ 
 equals the sum of the transmission rates.
 In Ref. \cite{previous-paper},
  the author derived approximate formulas
 to clarify the direct relationship between the transmission rate and
  the interlayer Hamiltonian.
The relation between the transmission rates
 and the dispersion lines was used in this derivation.
 The energy $E$, Hamiltonian $H$, and 
 overlap matrix $B$ were common to
  the exact calculations and the approximate formulas.
The approximate formulas indicated that
  the disagreement comes from differences in the interlayer Hamiltonian.
 The interlayer Hamiltonian is also important in multilayer graphenes  \cite{few-layer-graphite}  with regard to, e.g., transmission through a boundary
  between monolayer and bilayer graphene. \cite{mono-double-junction}
The TB Hamiltonian was used
in Ref. \cite{previous-paper}, but is insufficient
 for the interpretation of the first-principles
 calculations.
 Thus, both the exact method and
 the approximate formulas
 were adapted to a first-principles calculation code
 with an atomic orbital (AO) basis.
 The effectiveness of the approximate formulas
 was judged from their agreement with
 exact calculations.

\section{initial geometric structure and site  index}
Geometric optimization
 was performed with the initial structure
 defined as follows.
 In the initial structure, cylindrical coordinates $(r,\theta,z)= (r^{\mu},\theta^{\mu}_{l,m},z^{\mu}_{j,m})$ of the carbon atoms in the tube $\mu=O$
 (outer tube) and $\mu=I$ (inner tube)  are represented by
\begin{equation}
\theta^{\mu}_{l,m}=\frac{\pi}{n_\mu}\left(
l+\frac{-5-(-1)^{l+m}}{6}\right)
\label{atom-pos-theta}
\end{equation}
\begin{equation}
z^{\mu}_{j,m}=0.5a(2j+m) .
\label{atom-pos-z}
\end{equation}
The notation of Eqs. (\ref{atom-pos-theta}) and (\ref{atom-pos-z})
 is the same as Ref. \cite{previous-paper} except
 that the integer index $j$ in Ref. \cite{previous-paper}
 is replaced by $2j+m$ here.
The radius of the tube $\mu$ is denoted as $r^{\mu}=\sqrt{3}an_{\mu}/(2\pi)$,
 with   integers  $n_O=10, n_I=5$
 and the lattice constant $a = \sqrt{3} \times 0.142 $  nm.
The unit cells are numbered $j$ along the tube axis, while
 the atoms in each unit cell are numbered
  $m=-1,0$ and  $l=1,2,\cdots,2n_{\mu}$.
 As the DWNT considered here has no mirror plane  parallel to the tube axis,
 symmetric and anti-symmetric states are hybridized
  by the interlayer interaction.
We use $\mu, l,j,m$ as site indexes
 in the following.
 We discuss three structures: 
 an un-telescoped DWNT with  {\it five} unit cells ,
 a TDWNT, 
 and an un-telescoped DWNT with {\it infinite} length.
These are denoted as [fi], [td], and [in].
 Although Eqs. (\ref{atom-pos-theta}) and (\ref{atom-pos-z})
 are common among [fi], [td], and [in],
  the range of $j$ at which the corresponding atoms exist is different,
 as
\begin{equation}
1 \leq  j   \leq  5  \;\;\;\;{\rm \;\; in\;\; [fi],}
\label{case(a)}
\end{equation}
\begin{eqnarray}
1 &\leq&  j   < \infty 
\;\; \;\; {\rm in\;\; tube\;\;} O {\rm \;\;of \;\;[td]\;\;}  \nonumber \\
- \infty &\leq & j   \leq  L \;\; {\rm in\;\; tube\;\;} I {\rm \;\;of \;\;[td]}
\label{case(c)}
\end{eqnarray}
and
\begin{equation}
 -\infty < j  < \infty   \;\;\;\;{\rm \;\;in \;\;[in]}.
\label{case(b)}
\end{equation}
 where $L$ denotes the number of unit cells in the DWNT of [td].
 In the initial structure of [fi], the direction of the C-H bond
 is the same as the corresponding C-C
 bond and the C-H  bond length is 0.11 nm.

\section{  building block procedure}
 The SIESTA code is used 
 as the first-principles calculation of  structure [fi].\cite{siesta,siesta2,note} 
 The geometry is optimized by the conjugate gradients method
 with  the convergence criteria  0.04 eV/Ang.
 The optimized and initial  structures
 are superimposed on Fig.   2.
 We can see that
  the initial structure is almost maintained
 after the optimization.
 The minimum, average and maximum  bond lengths
 are 0.1175, 0.1177, 0.1181 nm for C-H
 and 0.1413, 0.1451, 0.1482 for C-C.
The output data after the geometrical optimization are
 used.
 Since the obtained Hamiltonian matrix has an atomic orbital (AO) basis,
 it can be divided into block matrices based on the unit cells.
The one-electron wave function $\psi$ is represented by 
 the linear combination of AOs  $ \{ \phi \}$,  as
\begin{equation}
\psi= \sum_{\mu=I,O}\sum_{j} \sum_{l=1}^{2n_{\mu}}\sum_{m=-1}^{0}
\sum_{\alpha}
d^{\mu}_{l,m,\alpha,j}
\phi^{\mu}_{l,m,\alpha,j} \;\;.
\label{Psi}
\end{equation}
In Eq. (\ref{Psi}), four single zeta AOs ($\alpha =$  2s, 2p$_x$, 2p$_y$, 2p$_z$) are considered per atom.

The secular  equations for structures [in] and [td]
 are represented by
\begin{equation}
 \sum_{\mu'=I,O}\sum_{j'}
H^{\mu,\mu' }_{j,j'}
 \vec{d}^{\;\mu'}_{j'} = E   \sum_{\mu'=I,O}\sum_{j'} B^{\mu,\mu' }_{j,j'}
 \vec{d}^{\;\mu'}_{j'}
\label{secular} 
\end{equation}
 where $H^{\mu,\mu'}_{j,j'}$  
 and $B^{\mu,\mu'}_{j,j'}$  
 are the block matrices of $H$ and $B$, respectively.
 The vector $\vec{d}^{\;\mu}_{j}$  is composed of $d^{\;\mu}_{l,m,\alpha,j}$ in Eq. (\ref{Psi}).
The ranges of $j'$ and $j$ in Eq. (\ref{secular})
 are shown in Eqs. (\ref{case(c)}) and (\ref{case(b)}).
From the obtained 
$H^{\mu,\mu',{\rm [fi]}  }_{j,j'}$ and $B^{\mu,\mu',{\rm [fi]}  }_{j,j'}$,
 the block matrices of structure [in] are defined 
 as
\begin{equation}
H^{\mu,\mu' }_{\eta ,\eta+\Delta j}
=\;^t\!H^{\mu',\mu }_{\eta+\Delta j,\eta}
 \equiv H^{\mu,\mu',[{\rm fi} ] }_{3,3+\Delta j}
\label{def-H}
\end{equation}
for $(\mu,\mu')=(I,I),(O,I), (I,O) $,
\begin{equation}
H^{O,O }_{\eta ,\eta+\Delta j}
=\;^t\!H^{O,O }_{\eta+\Delta j,\eta}
 \equiv H^{O,O,[{\rm fi} ] }_{3,3+\Delta j}
+\Delta \varepsilon  B^{O,O,[{\rm fi} ] }_{3,3+\Delta j}
\label{def-H2}
\end{equation}
and
\begin{equation}
B^{\mu,\mu'}_{\eta ,\eta+\Delta j}
=\;^t\!B^{\mu',\mu }_{\eta+\Delta j,\eta}
 \equiv B^{\mu,\mu',[{\rm fi}] }_{3,3+\Delta j}
\label{def-B}
\end{equation}
 where  $\eta$ is an arbitrary integer and $\Delta j=0,1,2$.
When $|\Delta j|>2$, $B^{\mu',\mu }_{\eta ,\eta+\Delta j}\equiv 0$
 and $H^{\mu',\mu,[\xi  ] }_{\eta ,\eta+\Delta j}\equiv 0$.
 Here we select block matrices $H^{\rm [fi]},B^{\rm [fi]}$
for which the boundary effect is relatively small and build them into the $H$ and $B$  matrixes of structure [in].
Once we obtain $H^{[{\rm fi} ]}$ and $B^{[{\rm fi} ] }$ from the SIESTA,  there is no adjustable parameter
 except for $\Delta \varepsilon$.
The purpose of introducing the parameter $\Delta \varepsilon$ 
 will be explained in Sec. IV.
 The $H$ and $B$  matrices of structure  [td] are  defined 
 also by the building block procedure as will be shown in Sec.VB.

\section{dispersion relation}
In order to discuss the dispersion relation of structure [in],
 it is useful to define
\begin{equation}
X^{D,D}
 \equiv
\left (
\begin{array}{cc}
X^{I,I}
& 
X^{I,O} \\
X^{O,I} & X^{O,O}
\end{array}
\right ) 
\label{def-DD}
\end{equation}
and
\begin{equation}
\widetilde{X}^{\mu',\mu}(k) \equiv \sum_{j=-M}^{M}
X^{\mu',\mu }_{0,j}e^{ikaj}
\label{def-X(k)}
\end{equation}
 where $X=H, B$.
The dispersion relation $E^{\mu}_{\tau}(k)$ 
and the wave function per unit cell $\vec{g}_{\tau}^{\mu}(k)$ 
 can be obtained by
\begin{equation}
\widetilde{H}^{\mu,\mu}(k) \vec{g}_{\tau}^{\mu}(k)
=E_{\tau}^{\mu}(k) \widetilde{B}^{\mu,\mu}(k) \vec{g}_{\tau}^{\mu}(k)
\label{Hg=EBg}
\end{equation}
 with the band index $\tau$.
According to Eqs. (\ref{def-DD}), (\ref{def-X(k)}) and (\ref{Hg=EBg}),
 $E^D_{\tau}(k)$ corresponds to a DWNT 
 {\it with} interlayer interaction,
 while   $E^I_{\tau}(k)$ and $E^O_{\tau}(k)$ correspond to
 {\it isolated} SWNTs.

The dispersion lines near the neutral Fermi level $E_F^{(0)}$ 
 are denoted by
  $E^{D}_{\pm 1}$, $E^{D}_{\pm 2}$, $E^{D}_{\pm 3}$, $E^{D}_{\pm 4}$
 for a DWNT with interlayer interaction,
 and
 $E^{O}_{\pm 1}$, $E^{I}_{\pm 1}$  $E^{O}_{\pm 2}$ and  $E^{I}_{\pm 2}$
 for isolated tubes $O$ and $I$.
 Figure 3 shows the dispersion lines numbered based on the index $\tau$ under the
 conditions $E_{|\tau|}^{\mu}\left( (-1)^\tau\frac{2\pi}{3a} \right)
 = E_{-|\tau|}^{\mu}\left(-(-1)^\tau\frac{2\pi}{3a} \right) \simeq E_F^{(0)}$
 and $ \tau \frac{dE_{\tau}^{\mu}}{dk}>0$.
For the isolated tubes $\mu=I$ and $\mu=O$,  $E^{\mu}_{\pm 1}$ 
 and $E^{\mu}_{\pm 2}$
 correspond to the symmetric and anti-symmetric states, respectively, with respect to the mirror plane parallel to the tube axis.
For a DWNT with interlayer interaction, however,
the mirror symmetry is broken
 and the band splitting is increased as  $E_3^D  -E_1^D > |E^O_1-E^I_1|$ and  
 $E_4^D  -E_2^D > |E^O_2-E^I_2|$.

 Figure 4 shows the dispersion relation of the 
  isolated SWNTs
  for $M=1$ (open circles)
 and $M=2$ (closed diamonds).
 The shift between $M=1$ and $M=2$
 is about 0.002 $\sim$ 0.003 Ry 
for the $E_{-1}$ band.
  In order to include this energy difference,
  the integer $M$ in Eq. (\ref{def-X(k)})
  is set to 2 hereafter.
 Squares and circles in Fig.   5
  show the intralayer and interlayer effective Hamiltonian
 elements $h(\vec{r},\vec{r} \;')$
, respectively, for the atomic distance 0.3 nm $ < |\vec{r}-\vec{r}\:'|
  < $ 0.45nm.
The effective Hamiltonian  $h(\vec{r},\vec{r}\:')$  is defined by
\begin{equation}
h \equiv \left|
\frac{
\sum_{\alpha,\alpha'} g^*_{\vec{r},\alpha}
H g_{\vec{r}\:',\alpha\;'}
}
{
\sqrt{\sum_{\alpha,\alpha'} g^*_{\vec{r},\alpha}
B g_{\vec{r},\alpha'}}
\sqrt{\sum_{\alpha,\alpha'} g^*_{\vec{r}\:',\alpha}
B g_{\vec{r}\:',\alpha'}}}
\right|
\label{def-effective-h}
 \end{equation}
where $g_{\vec{r},\alpha} $ denotes the component
 of $\vec{g}^I_{-1}, \vec{g}_{-1}^O$
 for $k=0.66\pi/a$, atomic position $\vec{r}$ and
 AO type $\alpha$=2s,2p$_x$,2p$_x$,2p$_z$.
 The spatial range of the effective Hamiltonian is considerably longer 
 compared to the TB Hamiltonian used in Ref. \cite{previous-paper}.
 Crosses represent the intralayer elements between the second nearest
 neighbor unit cells. They cause
 the energy difference  between $M=1$ and $M=2$  in Fig.   4.

 According to Ref.\cite{previous-paper}, we define the
 intrinsic shift as
\begin{equation}
\varepsilon_{\tau}
\equiv E^O_{\tau}(k)-E^I_{\tau}(k)\;\;.
\label{def-epsilon}
\end{equation}
Figure 4 shows a reasonable value of $\frac{dE^{\mu}}{dk}$.
 However, $|\varepsilon_{\tau}(k)|$ 
 is considerably larger
 than other first-principles calculation results,
 probably because of the finite size effect of structure [fi]
 used in the first step of the  building block procedure.
 The parameter $\Delta \varepsilon$ in Eq. (\ref{def-H})
 changes  $\varepsilon_{\tau} $ as
\begin{equation}
\varepsilon_{\tau}
= \varepsilon_{\tau}^{(0)}
+\Delta\varepsilon
 \label{epsilon-2}
\end{equation}
 where $\varepsilon_{\tau}^{(0)}$ denotes the intrinsic shift
 in Fig.   4.
 As  $\Delta\varepsilon$ is common to
 the exact and approximate methods,
 it is {\it not} a  parameter that can be adjusted to produce agreement between the results obtained by the two methods.
Figure 4  shows that $\varepsilon_{-1}^{(0)}=-0.022$  Ry
 and $\varepsilon_{2}^{(0)}=-0.028$  Ry at  $k=0.65\pi/a$.

When $\mu=D$, Eq. (\ref{Hg=EBg}) 
 can be transformed to
\begin{equation}
^t\vec{g}^{\;D}_{\tau'}(k)^* \left[\widetilde{H}^{D,D}(k)-E_{\tau'}^{D}(k) 
\widetilde{B}^{D,D}(k) 
\right] \vec{g}_{\tau'}^{D}(k)=0\;\;.
\label{derivation-ED-1}
\end{equation}
In the following, we consider the wavenumber $k$ 
 as a function of the energy $E$.
The wavenumbers of the {\it isolate} tubes $I$ and $O$
 are denoted by $k^I(E)$ and $k^O(E)$, respectively.
Neglecting  mixing between different $\tau$ values, 
the lowest order approximation is represented by
\begin{equation}
^t\vec{g}^{\;D}_{\tau'}(k) 
\simeq  (y_I \,^t\vec{g}^{\;I}_{\tau}(k^I),\,y_O \,^t\vec{g}^{\;O}_{\tau}(k^O) )
\label{assume}
\end{equation}
  where $\tau'=\tau, \tau+2\frac{\tau}{|\tau|}$.
Here the three different wavenumbers have the common energy as
$E=E^D(k)=E^I(k^I)=E^O(k^O)$ and $k \neq k^I \neq k^O $.

Using Eq. (\ref{assume}) and  

\begin{equation}
\widetilde{X}^{\mu',\mu}(k) \simeq \widetilde{X}^{\mu',\mu}(k^{\mu})+(k-k^{\mu})
\left. \frac{d}{dk}\widetilde{X}^{\mu',\mu} \right|_{k=k^{\mu}},
\label{deri-X}
\end{equation}

Eq. (\ref{derivation-ED-1}) can be approximated by
\begin{equation}
{\rm Re} [(A_{\tau}+C_{\tau})y_O^*y_I]
+ \sum_{\mu=I,O}
\frac{k-k^{\mu}_{\tau}}{\sqrt{b^I_{\tau}b^O_{\tau}}}
b^{\mu}_{\tau}
\frac{dE^{\mu}_{\tau}}{dk}
|y_{\mu}|^2 
\simeq 0
\label{derivation-ED-2}
\end{equation}
where
\begin{equation}
b_{\tau}^{\mu}\equiv\;
^{t}\!\vec{g}_{\tau}^{\mu}(k^{\mu})^*  
\widetilde{B}^{\mu,\mu}(k^{\mu}) \vec{g}_{\tau}^{\mu}(k^{\mu})\;\;.
\label{def-b}
\end{equation}

\begin{equation}
b_{\tau}^{\mu}\frac{ dE_{\tau}^{\mu}}{dk}
=
\left.\left( ^t\!\vec{g}_{\tau}^{\mu *}
\frac{d \widetilde{P}^{\mu,\mu}(k)}{dk}
 \vec{g}_{\tau}^{\mu}
\right)\right|_{k=k^{\mu}}
\label{2-app-2-2}
\end{equation}

\begin{equation}
A_{\tau}\equiv
\frac{
2\,^{t}\vec{g}_{\tau}^{\;O}(k^O)^*  
\widetilde{P}^{O,I}(k^I) \vec{g}_{\tau}^{\;I}(k^I)
}
{
\sqrt{b_{\tau}^{O}b_{\tau}^{I}}}
\label{def-A}
\end{equation}
\begin{equation}
\label{def-A'}
C_{\tau}
\equiv 
\frac{
2(k-k^I)}
{
\sqrt{b_{\tau}^{O}b_{\tau}^{I}}}
\left.\,^{t}\vec{g}_{\tau}^{\;O*}(k^O_{\tau})
\left(\frac{d\widetilde{P}^{O,I} }{dk} \vec{g}_{\tau}^{\;I}
\right) \right|_{k=k^{I}}
\end{equation}

\begin{equation}
\widetilde{P}^{\mu,\mu'}(k)
\equiv \widetilde{H}^{\mu,\mu'} (k)-E\widetilde{B}^{\mu,\mu'}(k)
\label{block-P-OI} 
\end{equation}

\begin{equation}
\frac{d \widetilde{P}^{\mu,\mu'}}{dk}
\equiv \frac{d \widetilde{H}^{\mu,\mu'}}{dk}-E\frac{d \widetilde{B}^{\mu,\mu'}}{dk} 
\end{equation}
Here Eq. (\ref{2-app-2-2}) will be proved in Appendix B.

From Eq. (\ref{derivation-ED-2}),   $k^D_{\tau'}$
  can be approximated by 
\begin{equation}
k^{D,A} _{\tau'}(E)
= \frac{1}{2}\left[
k^I_{\tau}(E)+k^O_{\tau}(E)
 \pm \Delta k_{\tau}^{D,A}
\right]
\label{approximate-ED}
\end{equation}
where 
\begin{equation}
\Delta k_{\tau}^{D,A}
 \equiv  
 |\Delta \widetilde{k}_{\tau}|\sqrt{1+x_{\tau}}
\label{deltak-def-D-Ap}
\end{equation}
\begin{equation}
\Delta \widetilde{k}_{\tau} \equiv k^O_{\tau}(E)-k^I_{\tau}(E)
\label{deltak-def-IO}
\end{equation}
and $x_{\tau}$ is a dimensionless parameter defined by
\begin{equation}
x_{\tau}
\equiv |A_{\tau}|^2\left|\frac{dk^I}{dE}\frac{dk^O}{dE}\right|(\Delta \widetilde{k})^{-2}.
\label{def-x}
\end{equation}
 The  band split along the $k$ axis 
 without interlayer interaction
 is denoted by Eq.(\ref{deltak-def-IO}),
 while that   with interlayer interaction
\begin{equation}
\Delta k^D_{\tau} \equiv \left|k^D_{\tau+2\frac{\tau}{|\tau|}}(E)- k^D_{\tau}(E)\right|
\end{equation}
is approximated by Eq.(\ref{deltak-def-D-Ap}).
Since $|C| \ll |A|$, the effect of Eq.(\ref{def-A'})  is neglected here.
Note that we define
  $\vec{g}^{O}_{\tau}$, $\vec{g}^{I}_{\tau}$,  $b^{O}_{\tau}$
$b^{I}_{\tau}$  , $k^{O}_{\tau}$
and  $k^{I}_{\tau}$ 
 {\it without} the interlayer matrixes $H^{O,I}$ and $B^{O,I}$.

\section{Transmission rate}
 In the following,
$T_{\tau,\tau'}$ denotes
the transmission rate from the $\tau'(= 1,2)$ channel of tube $I$
 to the $\tau(=1,2)$ channel of tube $O$. Here, the notation of $\tau$ 
 is common with that of Sec. IV, i.e., $\tau=1$ and $\tau=2$  are symmetric and anti-symmetric channels, respectively.

Since the interlayer Hamiltonian
 breaks the mirror symmetry,
 the inter-channel transmission rates
 $T_{1,2}$ and $T_{2,1}$ are not
zero. Nevertheless, the large difference 
 between $k_1 \simeq -2\pi/(3a) $ and $k_2 \simeq 2\pi/(3a)$
 suppresses $T_{1,2}$ and $T_{2,1}$.
 In the approximate formulas,
 we consider only the intra-channel transmission rates $T_{1,1}$ and $T_{2,2}$.

\subsection{Approximate formula}\label{sec-1}
 In order to clarify the relation
 of Eq. (\ref{def-A})
 to the effective interlayer interaction of Ref. \cite{previous-paper},
 we transform  Eq. (\ref{def-A})
 into
 \begin{equation}
 A = \frac{2\int \chi^{O}(1)^* (\widehat{H}-E)  \Psi^I(k)d^3\vec{r}}
{\sqrt{\int \chi^{O}(1)^*  \Psi^O(k)d^3\vec{r}}\sqrt{\int \chi^{I}(1)^*  \Psi^I(k)d^3\vec{r}}}
 \label{conclu-A} 
\end{equation}
 where $\widehat{H}$ is the Hamiltonian operator,
 \begin{equation}
 \Psi^{\mu}(k) \equiv \sum_{j=-\infty}^{\infty}e^{ikaj}\chi^{\mu}(j)\;,
 \label{conclu-Psi} 
 \end{equation}
 \begin{equation}
\chi^{\mu}(j) \equiv 
\sum_{l=1}^{2n_{\mu}}\sum_{m=-1}^{0}
\sum_{\alpha}
g^{\mu}_{l,m,\alpha,j}(k)\phi^{\mu}_{l,m,\alpha,j}
\label{conclu-chi}
 \end{equation}
and 
 $g^{\mu}_{l,m,\alpha,j}(k)$ denotes 
 the component of $\vec{g}^{\mu}(k)$.
 Equation (\ref{conclu-A})
 clearly indicates that
 $\frac{A}{2}$  equals
 the matrix element of the perturbation $H^{O,I}-EB^{O,I}$ between
 the {\it unperturbed} states $\Psi^O$ and $\Psi^I$
  per unit cell.
 When $\mu=O, I$,  Eq. (\ref{conclu-Psi}) is the eigenfunction of 
 the intralayer Hamiltonian $H^{\mu,\mu}$
 and bears no relation to the interlayer Hamiltonian $H^{O,I}$.
 Since $A$ of Ref. \cite{previous-paper}
 can also be represented by Eq. (\ref{conclu-A}),  
 the  physical meaning  of $A$ is common to
 the present paper and  Ref. \cite{previous-paper}.
 There is, however, another possible definition of $A$ where
  $\widetilde{P}^{O,I}=\widetilde{H}^{O,I}-E\widetilde{B}^{O,I}
$  is replaced by $\widetilde{H}^{O,I}$.
 The two definitions cause no difference when the interlayer
 overlap matrix $B^{O,I}$ is absent.
 Though the effect of $E\widetilde{B}^{O,I}$
 is also investigated in Sec.VI,  the definition (\ref{def-A})
 is chosen here.

The approximate formula of Ref.  \cite{previous-paper}
 is represented by
\begin{equation}
T_{\tau,\tau}'=\frac{|A_{\tau}|^{2}}{\varepsilon_{\tau}^2+|A_{\tau}|^{2}%
}\sin^{2}\left[ 
\sqrt{\varepsilon_{\tau}^2+|A_{\tau}|^{2}} \frac{dk}{dE}
L\frac{a}{2}
\right]  
\label{previous-T}
\end{equation}
where
\begin{equation}
 \frac{dk}{dE}
\equiv 
 \sqrt{\left|\frac{dk^O_{\tau}}{dE}\frac{dk^I_{\tau}}{dE}\right|}
\label{def-dE/dk}
\end{equation}
 is proportional to the geometrical mean of  the density of states.
Since the dispersion lines are almost straight and parallel,
\begin{equation}
\varepsilon_{\tau} \simeq
 \left(\frac{dk}{dE}\right)^{-1}  \Delta \widetilde{k}_{\tau}\;\;.
\label{def-epsilon2}
\end{equation}
Using Eqs. (\ref{deltak-def-D-Ap}), (\ref{def-x}), (\ref{def-dE/dk})
 and (\ref{def-epsilon2}), 
 we can transform  Eq.(\ref{previous-T}) to
\begin{eqnarray}
T_{\tau,\tau}' &=&
\left(1-\left| \frac{\Delta \widetilde{k}_{\tau}
}{\Delta k_{\tau}^{D,A}}
\right|^2\right)
\sin^{2}\left(  \frac{\Delta k^{D,A}_{\tau}}{2}La
\right)\nonumber \\
&=& 
 \frac{x_{\tau}}{1+x_{\tau}}
 \sin^{2}\left(  \sqrt{1+x_{\tau}} \frac{\Delta \widetilde{k}_{\tau}
}{2}La \right)
\label{T'}
\end{eqnarray}

\subsection{Exact numerical calculations}
 The scattering matrix method was used
as an exact method.\cite{S-matrix,Ando,tamura-method}
 We can also use the Green's function method, 
 which is useful for including inelastic scattering,
 electron correlation and finite bias.\cite{inelastic-Green,finite-bias}
 Without  these effects, however,
 the same transmission rate
 is obtained by both methods.\cite{Datta} 
 There are advantages of using
 scattering matrix method compared to the
 Green's function method.
 One is the explicit relation to the wave function,
 and the other is that we can estimate
 the numerical error from the unitarity of
 the scattering matrix.

We can obtain only
 the propagating states from Eq. (\ref{Hg=EBg}),
 but the evanescent wave states
 are also necessary for the exact calculation
 of the transmission rate.
In order to obtain both
 the propagating and evanescent states,
 the transfer matrix $\Gamma_{\mu}$ is defined
 as
\begin{equation}
\Gamma_{\mu} \equiv
\left(
\begin{array}
[c]{cccc}%
0,& 1, & 0, & 0\\
0,& 0, & 1, & 0\\
0, & 0, & 0, & 1\\
\widetilde{Y}_{2}^{\mu}, & \widetilde{Y}_{1}^{\mu}, & Y_{0}^{\mu}, & Y_{1}^{\mu}%
\end{array}
\right)  
\label{T-matrix}
\end{equation}
where
\begin{equation}
 Y_{j}^{\mu} \equiv -\left(  P_{3,5}^{\mu,\mu}\right)  ^{-1}
P_{3,3+j}^{\mu,\mu}
\label{def-Y}
\end{equation}
\begin{equation}
 \widetilde{Y}_{j}^{\mu} \equiv -\left(  P_{3,5}^{\mu,\mu}\right)^{-1}\:
^tP_{3,3+j}^{\mu,\mu}\;\;.
\label{def-Y2}
\end{equation}
The matrix $P$  is defined as
\begin{equation}
P^{\mu',\mu }_{j',j}
\equiv H^{\mu',\mu,[{\rm fi} ] }_{j',j}
-(E+\Delta \varepsilon \delta_{\mu',\mu}\delta_{\mu,O})B^{\mu',\mu, [{\rm fi} ] }_{j',j} \;\;.
\label{block-P} 
\end{equation}
Figure 6 shows the relation between  structure [fi] 
 and the  matrixes $P$.
 The matrix $P^{D,D}$ is composed of $P^{I,I},P^{O,I},P^{I,O}$ and $P^{O,O}$
 in the same way as Eq. (\ref{def-DD}).

The secular equation is represented by $\vec{e}_{j+1}=\Gamma_{\mu} \vec{e}_j$  where
$
 ^t\!\vec{e}_j \equiv (^t\!\vec{d}_{j-2}^{\;\mu},\,^t\!
\vec{d}_{j-1}^{\;\mu},\,^t\!
\vec{d}_{j}^{\;\mu},\,^t\!
\vec{d}_{j+1}^{\;\mu})$ 
 with $\mu$ corresponding to $j$
 as  $(\mu=I, j\leq -2)$, $(\mu=D, 3\leq j \leq L-2)$
 and $(\mu=O, L+3 \leq j)$.
For the propagating waves, the eigenvector $\vec{u}$
 and the eigenvalue $\lambda$ of the transfer matrix $\Gamma_{\mu}$ 
 satisfying $\Gamma_{\mu}\vec{u}= \lambda \vec{u}$  and $\vec{e}_j=\lambda^j\vec{u}$  can be related to $k$ and $\vec{g}^{\mu}$ of
 Eq. (\ref{Hg=EBg})
 as  $\lambda =e^{ika}$ and $\;^t\vec{u}=
\left(\,^{t}\!\vec{g}^{\mu} ,\;
\lambda\,^{t}\!\vec{g}^{\mu},\;\lambda^{2}\,^{t}\!\vec{g}^{\mu},\;
  \lambda^{3}\,^{t}\!\vec{g}^{\mu}\right)$.
 In the following, the dimension of $\vec{g}_{\tau}^{\;\mu}$ is denoted by $N_{\mu}$
 $(N_I=80, N_O=160, N_D=240 )$ .
The $4N_{\mu}$  independent eigenvectors of $4N_{\mu} \times
 4N_{\mu}$ matrix  $\Gamma_{\mu}$ 
are labeled by $\tau$  as follows.
 For the propagating waves   ($|\lambda_{\tau}^{\mu}|=1$ ),
 $\tau= \pm 1, \pm 2,\cdots,\pm N_c^{\mu}$.
For the  evanescent waves 
($|\lambda_{\tau}^{\mu}| \neq 1$ ),
 $ \tau= \pm (N_c ^{\mu}+1),\pm (N_c ^{\mu}+2),\cdots, \pm 2N_{\mu}$.
The sign of $\tau$ is chosen to
 be consistent with the propagation direction and decay direction
 when $|\tau| \leq N_c^{\mu}$ and $|\tau| \geq N_c^{\mu}+1$, respectively.
Here, the channel number $N_c^{\mu}$ denotes
 half the number of independent propagating waves
 in region $\mu$.
 In the present paper, 
 the energy $E$ is close to
 the neutral  Fermi level, 
 and thus $N_c^{I}=N_c^O=2$ and $N_c^{D}=4$.

The amplitude vector $\vec{d}_j$ is represented
 by 
\begin{equation}
 \vec{d}^{\;\mu}_j
=\sum_{\tau=-N_{\mu}}^{N_{\mu}}
\left( \lambda^{\;\mu}_{\tau}\right)^{j}\vec{g}^{\;\mu}_{\tau}
\gamma^{\;\mu}_{\tau}
\label{numerical-T-dj}
\end{equation}
where $\tau \neq 0$ and the correspondence between $\mu$ and $j$
 is represented by
 $(\mu=I, j\leq 0)$, $(\mu=D, 1\leq j \leq L)$
 and $(\mu=O, L+1 \leq j)$.

 In the periodic region  $(\mu=I, j\leq -2)$, $(\mu=D, 3\leq j \leq L-2)$
 and $(\mu=O, L+3 \leq j)$,
 the secular equations can be represented
 by
\begin{equation}
\vec{d}_{j+2}^{\;\mu}=\sum_{\Delta j=-2}^{-1}
 \widetilde{Y}^{\mu}_{|\Delta j|}\vec{d}_{j+\Delta j}^{\;\mu}+
\sum_{\Delta j=0}^{1}Y^{\mu}_{\Delta j}\vec{d}_{j+\Delta j}^{\;\mu}
\label{III}
\end{equation}
 with Eqs. (\ref{def-Y}) and (\ref{def-Y2}).
 In the transition region,   the secular equations can be represented
 by 
\begin{equation}
^tQ^I_{1}\vec{c}_{-3}^{\;I}+
Q^I_{0}\vec{c}_{-1}^{\;I}+
q_{ID}\vec{c}_{1}^{\;D}=0  
\label{IID}
\end{equation}
\begin{equation}
^tq_{ID} \vec{c}_{-1}^{\;I}+
q_0'\vec{c}_{1}^{\;D}+
Q^D_{1}\vec{c}_{3}^{\;D}=0 
\label{IDD}
\end{equation}
\begin{equation}
^tQ^D_{1}\vec{c}_{L-3}^{\;D}+
q_0''\vec{c}_{L-1}^{\;D}+
q_{DO}\vec{c}_{L+1}^{\;O}=0
\label{DDO}
\end{equation}
\begin{equation}
^tq_{DO}\vec{c}_{L-1}^{\;D}+
Q^O_{0}\vec{c}_{L+1}^{\;O}+
Q^O_{1}\vec{c}_{L+3}^{\;O}=0   \;\;.
\label{DOO} 
\end{equation}
where the matrixes $Q$  and the vector $\vec{c}$ 
are defined as
\begin{equation}
\vec{c}^{\;\mu}_j
\equiv
\left (
\begin{array}{c}
\vec{d}^{\;\mu}_{j} \\
\vec{d}^{\;\mu}_{j+1}
\end{array}
\right ) \;\;,
\label{def-c}
\end{equation}
\begin{equation}
Q^{\mu}_{0}
\equiv
\left (
\begin{array}{cc}
P^{\mu,\mu}_{3,3},& P^{\mu,\mu}_{3,4}  \\
^tP^{\mu,\mu}_{3,4},& P^{\mu,\mu}_{3,3}
\end{array}
\right ) \;\;,
\label{def-Q0}
\end{equation}
\begin{equation}
Q^{\mu}_1
\equiv
\left (
\begin{array}{cc}
P^{\mu,\mu}_{3,5},&  0  \\
P^{\mu,\mu}_{3,4},& P^{\mu,\mu}_{3,5 }
\end{array}
\right ) 
\label{def-Q1}
\end{equation}
and the matrixes $q$ are defined as
\begin{equation}
q_0'
\equiv
\left (
\begin{array}{cc}
P^{D,D}_{1,1},& P^{D,D}_{1,2}  \\
^tP^{D,D}_{1,2},& P^{D,D}_{2,2}
\end{array}
\right ) 
\label{def-q0-2}
\end{equation}
\begin{equation}
q_{ID}\equiv
\left (
\begin{array}{cccc}
P^{I,I}_{3,5}, & P^{I,O}_{3,5}, & 0, & 0\\
P^{I,I}_{3,4}, 
& 
P^{I,O}_{3,4}, & P^{I,I}_{3,5}, & P^{I,O}_{3,5}
\end{array}
\right ) 
\label{def-qID}
\end{equation}
\begin{equation}
q_0''
\equiv
\left (
\begin{array}{cc}
P^{D,D}_{4,4},& P^{D,D}_{4,5}  \\
^tP^{D,D}_{4,5},& P^{D,D}_{5,5}
\end{array}
\right ) 
\label{def-q0-3}
\end{equation}
\begin{equation}
^tq_{DO}\equiv
\left (
\begin{array}{cccc}
^t\!P^{I,O}_{3,5}, & ^t\!P^{O,O}_{3,5}, &  ^t\!P^{I,O}_{3,4},  & ^t\!P^{O,O}_{3,4} \\
0 &0 & ^t\!P^{I,O}_{3,5}, & ^t\!P^{O,O}_{3,5} 
\end{array}
\right ) \;\;.
\label{def-qDO}
\end{equation}

Although the coefficients $\{ \gamma_{\tau}^I \}$, $\{ \gamma_{\tau}^O \}$
 and $\{ \gamma_{\tau}^D \}$
  in Eq. (\ref{numerical-T-dj})
 can be chosen  arbitrarily
 to satisfy Eq. (\ref{III}),
 they are subject to the conditions expressed in 
Eqs. (\ref{IID}),(\ref{IDD}),(\ref{DDO}) and ,(\ref{DOO}).
 Using 
Eqs. (\ref{numerical-T-dj}),(\ref{IID}),(\ref{IDD}) and (\ref{def-c}),  we can obtain a
 $ 320\times 320$
 matrix $\sigma^{ID}$ 
satisfying
$(^t \vec{\gamma}^{\;I}_{-},^t \vec{\gamma}^{\;D}_{+} )
=
(^t \vec{\gamma}^{\;I}_{+},^t \vec{\gamma}^{\;D}_{-} )\,^t\sigma^{ID}
$, 
where  $ ^t \vec{\gamma}^{\mu}_{\pm } 
=(\gamma^{\mu}_{\pm 1},\gamma^{\mu}_{\pm 2},\cdots,
\gamma^{\mu}_{\pm N_{\mu}})$. The matrix $\sigma^{DO}$ 
satisfying
$(^t \vec{\gamma}^{\;D}_{-},^t \vec{\gamma}^{\;O}_{+} )
=(^t \vec{\gamma}^{\;D}_{+},^t \vec{\gamma}^{\;O}_{-} )
\,^t\sigma^{DO}$ can be  obtained in the same way.
By eliminating $\vec{\gamma}^{\;D}_{\pm}$ from these equations,
we can obtain the  matrix
 $\sigma^{IO}$
satisfying
$(^t \vec{\gamma}^{\;I}_{-},^t \vec{\gamma}^{\;O}_{+} )
=(^t \vec{\gamma}^{\;I}_{+},^t \vec{\gamma}^{\;O}_{-} )
^t\sigma^{IO}$.
The $4 \times 4$  scattering matrix $S^{IO}$  is obtained
 from the $240 \times 240$  matrix $\sigma^{IO}$
 as $(S^{IO})_{\tau,\tau'}=(\sigma^{IO})_{\tau,\tau'},\;
S^{IO}_{\tau+2,\tau'}=\sigma^{IO}_{\tau+80,\tau'},\;
S^{IO}_{\tau,\tau'+2}=\sigma^{IO}_{\tau,\tau'+80}$ and $S^{IO}_{\tau+2,\tau'+2}=\sigma^{IO}_{\tau+80,\tau'+80}$,
 where $\tau=1,2$ and $\tau'=1,2$.
From the scattering matrix, we can obtain $T_{\tau,\tau'}=|S_{\tau+2,\tau'}^{I	O}|^2$. 

The difference between $^tS^{IO*}S^{IO}
$ and the unit matrix
 is represented by
\begin{equation}
{\rm err} (S^{IO})\equiv \sum_{l_1=1}^4\sum_{l_2=1}^4
\left|\left(\sum_{l_3=1}^4S^{IO*}_{l_3,l_1}S^{IO}_{l_3,l_2} \right) 
-\delta_{l_1,l_2}\right|^2\;.
\label{err(S)}
\end{equation}
 When there is no numerical error,
  $S^{IO}$ must be unitary 
  and Eq. (\ref{err(S)})  must be zero
 with the normalization shown in Appendix A. \cite{Datta,S-matrix-Buttiker}
 As the unitarity represents
 that the sum of transmission rate and reflection rate
 must be unity,  the  numerical  error in the transmission rate
 is estimated to be Eq. (\ref{err(S)}).

\section{results}
 Figure 7 shows the dispersion lines 
  for   (a) $\Delta\varepsilon = -0.015$ , (b) $\Delta\varepsilon =$ 0, and (c)  $\Delta\varepsilon =$ 0.020  Ry.
The relationship between $\Delta\varepsilon$ and $\varepsilon_{\tau}$
 is shown by Eq. (\ref{epsilon-2}).
 At $k=0.65\pi/a$,
  $\varepsilon_{-1}= -0.022+\Delta\varepsilon$
 and
  $\varepsilon_{2}= -0.028+\Delta\varepsilon$.
 Crosses  represent  $E^O_{\tau}(k)$ and $E^I_{\tau}(k)$ {\it without}  interlayer interaction.
In Fig.   7 (c),  $E_{-1}^O$ and  $E_{-1}^I$ almost coincide with each other
 as $\varepsilon_{-1} \simeq 0$.
The dispersion lines {\it with} interlayer interaction
 are shown by circles  for the exact $E^D(k)$ and  by solid lines  for the approximate  $E^{D,A}(k)$   defined by Eqs. (\ref{approximate-ED}), (\ref{deltak-def-D-Ap}), (\ref{deltak-def-IO}) and (\ref{def-x}).
 Since $|A_2| \ll |\varepsilon_2|$,
 $E_2^O$ and $E_2^I$ 
 are almost the same
 as  $E_2^{D,A}$ and $E^{D,A}_4$, respectively.
In the symmetric bands specified by $\tau =-1$,  on the other hand, the relatively large 
$|A_{-1}|$ causes a
 clear difference between the solid lines
 and the crosses.
As $|A_{-1}|$ is almost constant, $E^I_{-1}$ and $E^O_{-1}$ 
 are almost parallel to $E^{D,A}_{-1}$ and 
$E^{D,A}_{-3}$. 
The solid lines reproduce the band split of $\tau=-1$ channel,
 but underestimate that of $\tau=2$ channel.

Figures  8 and 9 shows  $T_{1,1}$ and $T_{2,2}$, respectively, 
 as a function of the number of unit cells $L$.
The circles  represent the exact transmission rate.
The values of $\Delta\varepsilon$ in Figs. 8 and 9 are the same as in Fig.   7.
Because $L$ must
  be larger than three in the exact calculation
  with  Eqs. (\ref{IDD}) and (\ref{DDO}), $L \geq 4$ in Figs. 8 and 9.
Triangles in Fig.  9 represent
 Eq. (\ref{err(S)})
 showing
 the numerical errors 
 of $T_{1,1}+T_{2,2}+T_{2,1}+T_{1,2}$ in the exact calculation.
 The average and maximum of   Eq. (\ref{err(S)})
 are
 $4.9\times 10^{-4}$ and $1.4\times 10^{-3}$.
The magnitude of $T_{1,2}+T_{2,1}$ obtained from the
 numerical calculation is also $10^{-4} \sim 10^{-3}$.
 This indicates that $T_{1,2}+T_{2,1}$ cannot be discussed
  with sufficient accuracy.
 Thus we focus our discussion
 on $T_{1,1}$ and $T_{2,2}$.
 Although  the interlayer Hamiltonian could
  cause a pseudo-gap 
 in the energy band
 of the DWNT,\cite{pseudo-gap}
 there is no pseudo-gap  at the energy  $E= -0.4$  Ry chosen here.
 The circles shows a rapid oscillation
 superposed on a slow oscillation.
 The rapid oscillation is due to a standing wave caused by a
 large intralayer reflection at the
 open edges $j=1,L$ in structure [td].
 As the wavenumber $k$  is close to $\pm 2\pi/(3a)$,
 the period of the rapid oscillation is close to three.
 In order to see the slow oscillation of the exact $T(L)$,  
\begin{equation}
T^{ave}(L) \equiv \frac{T(L-1)+T(L)+T(L+1)}{3}
\label{T-ave} 
\end{equation}
 is shown by the dot-dashed lines.
In order to estimate the edge effect, $q_0'$ in Eq. (\ref{IDD})
 and $q_0''$ in Eq. (\ref{DDO}) are replaced by $Q_0^D$.
 The exact  $T_{1,1}$  and  $T_{1,1}^{ave}$ 
 with this replacement are shown by crosses and dashed lines, respectively,
 in Fig.   8.  Agreement between the dashed line and dot-dashed line
 indicates that the slow oscillation defined by 
 (\ref{T-ave}) is not sensitive to the detail of the edge modeling.
Though crosses and dashed lines are omitted in Fig. 9,
 it is also confirmed that the two edge models show similar $T_{2,2}^{ave}$.

 The thick solid  lines represent
 $T'$  defined by Eq. (\ref{T'}),
 while the thin solid lines represent Eq.(\ref{T'})
 where $E\widetilde{B}^{O,I}$ in  Eq.(\ref{def-A}) is replaced by zero,
 i.e.,  the interlayer overlap matrix is set to zero.
 The thin solid lines are shown only for the first period.
 Though the thin solid lines are close to $T^{ave}$
 for the small $L$, the thick solid lines  qualitatively
 reproduce $T^{ave}$ for the wider range of $L$.
 Thus we concentrate our discussion to the thick
 solid lines below.
 The effective interlayer interaction
 of the symmetric state $A_{1}$
 is much larger than that of the anti-symmetric state
 $A_2$. This is the reason why $T_{1,1} \gg T_{2,2}$.
In the derivation of Eq. (\ref{previous-T})
 shown
 in Ref. \cite{previous-paper}, 
  the  interlayer Hamiltonian 
 is the same as  that for structure [td],
 while
the intralayer Hamiltonian
 is the same as  that for structure [in].
 Thus,  Eq. (\ref{previous-T}) does not include  the strong intralayer reflection at the edge $j=1,L$ of  structure [td].
This is why
the rapid oscillation does not appear in Eq. (\ref{T'}). 

 The period of the exact $T^{ave}$ is close to $2\pi/\Delta k^{D}$
 where $\Delta k_{\tau}^D$ denotes the {\it exact} band split of 
 channel $\tau$ with interlayer interaction.
 Since $\Delta k_1^{D,A} \simeq \Delta k_1^D$,
 the thick solid line  and the dot-dashed line
 have almost the same period in Fig. 8.
 On the other hand,
the thick solid lines show
 the longer period than the dot-dashed lines
 in Fig. 9 because $\Delta k_2^{D,A} < \Delta k_2^D$.
Nevertheless the peak height of $T^{ave}$ 
 is qualitatively reproduced by $x/(1+x)$ of Eq. (\ref{T'}) 
 both in Fig. 8 and in Fig. 9.

\section{discussion}
 In the present paper,
 we consider the differential conductance $dI/dV$ 
 at zero bias voltage $V=0$.
 We have to use more sophisticated numerical
 methods to obtain the effect of the finite $V$ 
 on the current $I$. \cite{finite-bias}
 If the rigid band picture is effective, however,
 the current $I$ for finite $V$ can be approximated  by 
 \begin{equation}
 I=\frac{e}{h}\int_{E_F}^{E_F+eV} T(E,\varepsilon+eV)dE
 \label{discussion-T}
 \end{equation}
 where $E_F$ is the Fermi level  of tube $O$
 and  the interlayer transmission rate 
 is represented by $T(E,\varepsilon)$  
 as a function of the energy $E$ and 
 the intrinsic shift $\varepsilon$.

 The effective interlayer interaction
 $A$ defined by Eqs. (\ref{def-A}) 
 enables us to analyze the effects of the interlayer
 Hamiltonian on the transmission rate.
 For example, we can discuss
 the difference between the TB used in Ref. \cite{previous-paper}
 and the SIESTA used here.
 The longer cut-off distance of the interlayer Hamiltonian
 decreases $|A_2|$ as was discussed in Ref. \cite{previous-paper}.
  The cut-off distance for SIESTA
 is much larger than that for TB as seen in Fig.   5.
 This is why  $|A_2/A_1|$ and $T_{2,2}$ for SIESTA
  are negligible compared to those for TB.
 Using the perturbation formulas,
 we can distinguish the effects 
 of the intralayer Hamiltonian
 $H^{O,O},H^{I,I}$ from
  those of the interlayer Hamiltonian
 $H^{O,I}$.
Rigorously speaking, the interlayer interaction
 influences
 $H^{I,I}$ and $H^{O,O}$ 
 in the self consistent calculation  of 
 structure [fi] for the  building block procedure.
 Nevertheless, 
  $H^{I,I}$ and $H^{O,O}$ are approximately considered
 as  the Hamiltonian of isolated SWNTs.
 As long as  a real space basis
 such as an AO basis is used, 
  the calculation methods in the present paper
  can be applied to  various other
 first-principles calculation codes.

 Equation (\ref{T'})
  is effective
 both  in the SIESTA Hamiltonian
 and  in the TB Hamiltonian.
We have to be careful that
 the effective interlayer interaction (\ref{def-A}) 
 must include the interlayer overlap matrix $B^{O,I}$
 that is absent in Ref.\cite{previous-paper}.
Using  Eq.(\ref{T'}), we can predict  the amplitude of $T^{ave}(L)$ as seen in Figs. 8 and 9. Though the precision of Eq.(\ref{T'}) is lower in Fig. 9,
 the qualitative dependence on $\varepsilon$ is reproduced.
 The period of $T_{1,1}^{ave}(L)$ is also reproduced well,
 but that of $T_{2,2}^{ave}$ is  overestimated.
 It is common to TB and SIESTA that Eq.(\ref{T'}) is more
  effective for $T_{1,1}$ than for $T_{2,2}$. 
The period $2\pi/\Delta k$ has been discussed
 in other papers, but  the amplitude
formula $ x_{\tau}/(1+x_{\tau})$
 in Eq.(\ref{T'})  is  proposed firstly here.
Note that the dimensionless parameter $x_{\tau}$ 
 is defined by Eq. (\ref{def-x}) without a fitting parameter.
Owing to the linear dispersion,
  the amplitude of $T_{1,1}'$ are represented
 by the energy ratios
 as \begin{equation}
1-\left| \frac{\Delta \widetilde{k}_{1}
}{\Delta k^D_{1}}
\right|^2
=1 -\left|\frac{\varepsilon_{1}}
{E^{D}_{3}-E^{D}_{1}}
\right|^2
\label{T''-amp}
\end{equation}
When $|\varepsilon_1|/\left( E_3^D-E_1^D \right) \simeq 1$
, Eq.(\ref{T''-amp}) is close to 
$2 \left( E_3^D-E_1^D-|\varepsilon_1|  \right)/|\varepsilon_1|$.
 Since $ E_3^D-E_1^D-|\varepsilon_1| $
 is the energy shift
 caused by the interlayer interaction,  its ratio to  $|\varepsilon_1|$ can be considered as the Thouless number.\cite{Thouless}
 Author expects that
 Eq. (\ref{T'})  represent  the generalized Thouless number analysis
 and  can be applied to other related systems
 with commensurate interlayer configurations.

\appendix
\section{Relation between the normalization and probability flow}

The probability flow $J(\gamma'|\gamma)$ 
from orbital $\gamma=(\mu,l,m,\alpha,j)$
 to orbital $\gamma'=(\mu',l',m',\alpha',j')$
 is represented by
\begin{equation}
J(\gamma'|\gamma) \equiv {\rm Im}\left[
(H_{\gamma',\gamma}-EB_{\gamma',\gamma})
d_{\gamma'}^*d_{\gamma}
\right]
\label{J-gamma}
\end{equation}
 where $H, B$ and $E$ represent the Hamiltonian matrix, the overlap matrix
 and the energy, respectively.
Equation (\ref{J-gamma}) satisfies the 
 two necessary conditions
 for  stationary flow: 'direction of
 flow' represented by  $J(\gamma|\gamma') = -J(\gamma'|\gamma) $
 and   'conservation of probability ' represented by
\begin{equation}
\sum_{\gamma} J(\gamma'|\gamma) =0\;\;.
\label{conservation-J}
\end{equation}
Equation (\ref{conservation-J}) is derived from the secular Equation  $\sum_{\gamma}(H_{\gamma',\gamma}-EB_{\gamma',\gamma})d_{\gamma}=0$.
Using Eq. (\ref{J-gamma}), the probability flow $I_{tot}(j)$  through the cross section at $z=(j-\frac{1}{2})a$ is
 represented by
\begin{equation}
I_{tot}(j) =
 I^{\mu}_{j,j-1}+
 I^{\mu}_{j+1,j-1}+
 I^{\mu}_{j,j-2}
\label{J-tot-def}
\end{equation}
where   
\begin{equation}
I^{\mu}_{j,j'} \equiv \sum_{l,m,\alpha}
\sum_{l',m',\alpha'}
 J(\mu,l,m,\alpha,j|\mu,l',m',\alpha',j')\;.
\label{J-tot-def2}
\end{equation}
Corresponding to $M=2$ in Eq. (\ref{def-X(k)}),
$I^{\mu}_{j,j'}=0$ when $|j-j'| \geq 3$.
Using Eqs. (\ref{block-P}), (\ref{numerical-T-dj}), (\ref{def-Q1}), 
 (\ref{J-gamma}) and (\ref{J-tot-def2}),  Eq. (\ref{J-tot-def})
 can be expanded as
\begin{equation}
I_{tot}(j) 
=
\sum\limits_{\tau=-N_{\mu}}^{N_{\mu}}\;
\sum\limits_{\tau^{\prime}=\tau}^{N_{\mu}}
\left (1-\frac{\delta_{\tau,\tau'}}{2}\right)\left[
\eta_{\tau^{\prime},\tau}^{\mu}+
\eta_{\tau,\tau^{\prime}}^{\mu}
\right]
\label{J-tot}
\end{equation}
 where $\tau \neq 0,\tau'\neq 0$,
\begin{equation}
\eta_{\tau^{\prime},\tau}^{\mu}  \equiv {\rm Im}\left[
\left(  
\lambda_{\tau^{\prime}}^{\mu\ast}\lambda_{\tau}^{\mu}\right)^j
  Z_{\tau^{\prime},\tau}^{\mu}\,\gamma_{\tau^{\prime}}^{\mu\ast
}\gamma_{\tau}^{\mu}\right]  
\label{eta}
\end{equation}
\begin{equation}
Z^{\mu}_{\tau^{\prime},\tau}\equiv 
\left(  \lambda_{\tau}^{\mu}\right)^{-2}\left(  ^{t}\vec{f}_{\tau'}^{\;\mu\ast}\right)  \,^tQ_{1}^{\mu}
 \vec{f}_{\tau}^{\;\mu} 
\label{def-Z}
\end{equation}
and
\begin{equation}
\vec{f}_{\tau}^{\;\mu} \equiv \left(
\begin{array}
[c]{c}%
\vec{g}_{\tau}^{\;\mu} \\
  \lambda_{\tau}^{\mu}\vec{g}_{\tau}^{\;\mu}%
\end{array}
\right)  \;\;.
\label{def-f}
\end{equation}
In Eq. (\ref{J-tot}), the correspondence between
$\mu$ and $j$
 is represented by
 $(\mu=I, j\leq -2)$, $(\mu=D, 3\leq j \leq L-2)$
 and $(\mu=O, L+3 \leq j)$.

On the other hand,
 we can obtain
\begin{equation}
\left((\lambda^{\mu}_{\tau})^{-2}\:^tQ^{\mu}_1
+ Q^{\mu}_0
+(\lambda^{\mu}_{\tau})^2 Q^{\mu}_1
\right)
\vec{f}_{\tau}^{\mu}=0
\label{appendix-Q}
\end{equation}
from the secular equations.
Multiplying Eq. (\ref{appendix-Q}) by 
$^{t}\vec{f}_{\tau^{\prime}}^{\;\mu*}$,
 we can obtain
\begin{equation}
\left(  \lambda_{\tau}^{\mu}\lambda_{\tau^{\prime}}^{\mu\ast}\right)
^2 Z_{\tau,\tau^{\prime}}^{\mu*}+Z_{\tau^{\prime},
\tau}^{\mu}+\,^{t}\vec{f}_{\tau^{\prime}}^{\;\mu\ast}\,Q_{0}^{\mu}\vec{f}_{\tau}^{\;\mu}=0\;\;.
\label{apenndix-Z} 
\end{equation}
Exchanging
 $\tau$ and
 $\tau^{\prime}$ in the complex conjugate of Eq. (\ref{apenndix-Z}),
 we can also obtain
\begin{equation}
\left(  \lambda_{\tau}^{\mu}\lambda_{\tau^{\prime}}^{\mu\ast}\right)^2
  Z_{\tau^{\prime},\tau}^{\mu}+Z_{\tau,\tau^{\prime}}^{\mu *
}
+
\,^{t}\vec{f}_{\tau^{\prime}}^{\;\mu\ast}\,Q_{0}^{\mu}\vec{f}_{\tau}^{\;\mu}=0
\;\;.
\label{apenndix-Z-2} 
\end{equation}
When the condition
\begin{equation}
(\lambda_{\tau}^{\mu}\lambda_{\tau^{\prime}}^{\mu\ast})^4 = 1
\label{appendix-condition}
\end{equation}
 does {\it not}  hold,
we can  show 
\begin{equation}
Z_{\tau^{\prime},\tau}^{\mu}=Z_{\tau,\tau'}^{\mu *}
=
\frac
{-\,^{t}\vec{f}_{\tau^{\prime}}^{\;\mu\ast}\,Q_{0}^{\mu}\vec{f}_{\tau}^{\;\mu}}
{ \left(  \lambda_{\tau}^{\mu}\lambda_{\tau^{\prime}}^{\mu\ast}\right)^2
+1}
\label{appendi
x-Z-Z}
\end{equation}
 from Eqs. (\ref{apenndix-Z}) and (\ref{apenndix-Z-2}) .
Thus,
 $\eta_{\tau^{\prime},\tau}^{\mu}+\eta_{\tau,\tau^{\prime}}^{\mu}$ 
  has a nonzero value only when  the condition (\ref{appendix-condition}) 
 holds. 
 Under the condition (\ref{appendix-condition}), we only have to consider  three cases
 (i) $|\lambda_{\tau}^{\mu}|=|\lambda_{\tau'}^{\mu}|^{-1} \neq 1$,
 (ii) $\tau=\tau', |\lambda_{\tau}^{\mu}|=1 $,
 and (iii)  $\tau \neq \tau', 
|\lambda_{\tau}^{\mu}|=|\lambda_{\tau'}^{\mu}|=1$.
When $|\lambda_{\tau_1}^O| >1$ and $|\lambda_{\tau_2}^I| < 1$, 
 $\gamma_{\tau_1}^O=\gamma_{\tau_2}^I=0$  because
 finite values of $\gamma_{\tau_1}^O$ and $\gamma_{\tau_2}^I$ 
 cause  the divergence of $|\vec{d}_j^{\;\mu}|$ at $j= \infty$
 and $j=-\infty$, respectively.
 Thus,  $\gamma_{\tau'}^{\mu*}\gamma_{\tau}^{\mu}$ in Eq. (\ref{eta})  must be zero in case (i).
In case (iii), 
$\left(\lambda_{\tau}^{\mu}\right)^4$
 and
$\left(\lambda_{\tau'}^{\mu}\right)^4$
 are accidentally degenerate.
Since this degeneracy is lifted by
 an infinitesimal change in the energy $E$,
 case (iii) can be excluded.
Since  $\eta_{\tau^{\prime},\tau}^{\mu}+\eta_{\tau,\tau^{\prime}}^{\mu}$  
 is nonzero only in case (ii),
 Eq. (\ref{J-tot}) is represented by
\begin{equation}
 I_{tot}(j)=\sum_{\tau=1}^{N_c^{\mu}}
{\rm Im}(Z^{\mu}_{\tau,\tau}) 
|\gamma^{\mu}_{\tau}|^2+{\rm Im}(Z^{\mu}_{-\tau,-\tau}) |\gamma^{\mu}_{-\tau}|^2\;\;.
\label{J-tot-gamma}
\end{equation}
Equations (\ref{J-tot}) and (\ref{eta})
do not depend on $j$ because
 $\lambda_{\tau^{\prime}}^{\mu\ast}\lambda_{\tau}^{\mu}=\lambda_{\tau}^{\mu\ast}\lambda_{\tau}^{\mu}= |\lambda_{\tau}^{\mu}|^2 =1$   in case (ii).

When the vectors $\vec{f}_{\tau}^{\;\mu}$  
 are normalized as ${\rm Im}(Z_{|\tau|,|\tau|})=1$
 and ${\rm Im}(Z_{-|\tau|,-|\tau|})=-1$,
 we can show
\begin{eqnarray}
I_{tot} &=& \sum_{\tau=1}^{N_c^{I}}(|\gamma_{\tau}^I|^2-
|\gamma_{-\tau}^I|^2)=
\sum_{\tau=1}^{N_c^{D}}(|\gamma_{\tau}^D|^2-
|\gamma_{-\tau}^D|^2)
\nonumber \\
&=&
\sum_{\tau=1}^{N_c^{O}}(|\gamma_{\tau}^O|^2-
|\gamma_{-\tau}^O|^2)\;.
\label{appendix-ggg}
\end{eqnarray}
 Here, conservation of probability
  guarantees that  $I_{tot}$ is  common to regions $I$, $D$, and $O$.
Eq. (\ref{appendix-ggg})
is equivalent to  $^tS^{ID*}S^{ID}=1,$ $^tS^{DO*}S^{DO}=1$ and $^tS^{IO*}S^{IO}=1$.

\section{relation between  the normalization and group velocity}
From Eqs. (\ref{Hg=EBg}) and (\ref{def-b}), we can obtain

\begin{equation}
b_{\tau}^{\mu}E_{\tau}^{\mu}
=\!^t \vec{g}_{\tau}^{\mu *}
\widetilde{H}^{\mu,\mu} \vec{g}_{\tau}^{\mu}\;.
\label{2-app-1}
\end{equation}

Differentiating Eq. (\ref{2-app-1}) with respect to $k$,
\begin{eqnarray}
b_{\tau}^{\mu}\frac{ dE_{\tau}^{\mu}}{dk}
&= &-
 \frac{db_{\tau}^{\mu}}{dk}E_{\tau}^{\mu}+
\frac{d \;^t\vec{g}_{\tau}^{\mu *}}{dk}
\widetilde{H}^{\mu,\mu} \vec{g}_{\tau}^{\mu}
 \nonumber \\
&&
+ ^t\!\vec{g}_{\tau}^{\mu *}
\frac{d\widetilde{H}^{\mu,\mu}}{dk}
 \vec{g}_{\tau}^{\mu}
+ \;^t\!\vec{g}_{\tau}^{\mu *}
\widetilde{H}^{\mu,\mu}
\frac{ d\vec{g}_{\tau}^{\mu}}{dk}\;\;.
\label{3-14-tuika} 
\end{eqnarray}
Using Eqs. (\ref{def-X(k)}),(\ref{Hg=EBg}),(\ref{def-b}) and (\ref{block-P}),
 Eq. (\ref{3-14-tuika}) is represented by

\begin{eqnarray}
b_{\tau}^{\mu}\frac{ dE_{\tau}^{\mu}}{dk}
&=&
^t\!\vec{g}_{\tau}^{\mu *}
\left[
\frac{d\widetilde{H}^{\mu,\mu}}{dk}
-E_{\tau}^{\mu}\frac{d\widetilde{B}^{\mu,\mu}}{dk}
\right]
 \vec{g}_{\tau}^{\mu}
 \nonumber \\
&=& 2{\rm Re} \left[
\:^t\!\vec{g}_{\tau}^{\mu *}
\sum_{\Delta j=1}^2 P^{\mu,\mu}_{3,3+\Delta j}\left(\frac{d
}{dk} e^{ikaj}\right)
 \vec{g}_{\tau}^{\mu} \right]\;\;.
\label{2-app-2}
\end{eqnarray}

From Eqs. (\ref{def-Q1}),(\ref{def-Z})
 and (\ref{def-f}),
Eq. (\ref{2-app-2}) is represented by
\begin{equation}
b_{\tau}^{\mu}\frac{ dE_{\tau}^{\mu}}{dk} =2a {\rm Im} \left(Z^{\mu}_{\tau,\tau}\right)\;\;.
\label{2-app-3}
\end{equation}

With  Eqs. (\ref{def-A}), (\ref{def-dE/dk}) and (\ref{2-app-3}), 
Eq. (\ref{def-x}) can be represented by
\begin{equation}
x_{\tau} =
\frac{
\left|^{t}\!\vec{g}_{\tau}^{\;O}(k_{\tau}^O)^*  
\widetilde{P}^{O,I}(k_{\tau}^I) \vec{g}_{\tau}^{\;I}(k_{\tau}^I)
\right|^2
}
{
\left|{\rm Im} \left(Z^{O}_{\tau,\tau}\right)
{\rm Im} \left(Z^{I}_{\tau,\tau}\right)
\right|
a^2\Delta \widetilde{k}_{\tau}^2
}\;.
\end{equation}

When the normalization 
$\left|{\rm Im} \left(Z^{O}_{\tau,\tau}\right)\right|
=\left|{\rm Im} \left(Z^{I}_{\tau,\tau}\right)
\right|=1$ is used,
\begin{equation}
x_{\tau}
=
\frac{
\left|^{t}\!
\vec{g}_{\tau}^{\;O}(k_{\tau}^O)^*  
\widetilde{P}^{O,I}(k_{\tau}^I) \vec{g}_{\tau}^{\;I}(k_{\tau}^I)
\right|^2
}
{
a^2(k_{\tau}^O-k_{\tau}^I)^2
}\;.
\end{equation}

\begin{figure}
  \epsfxsize=0.5\columnwidth
\centerline{\hbox{   \epsffile{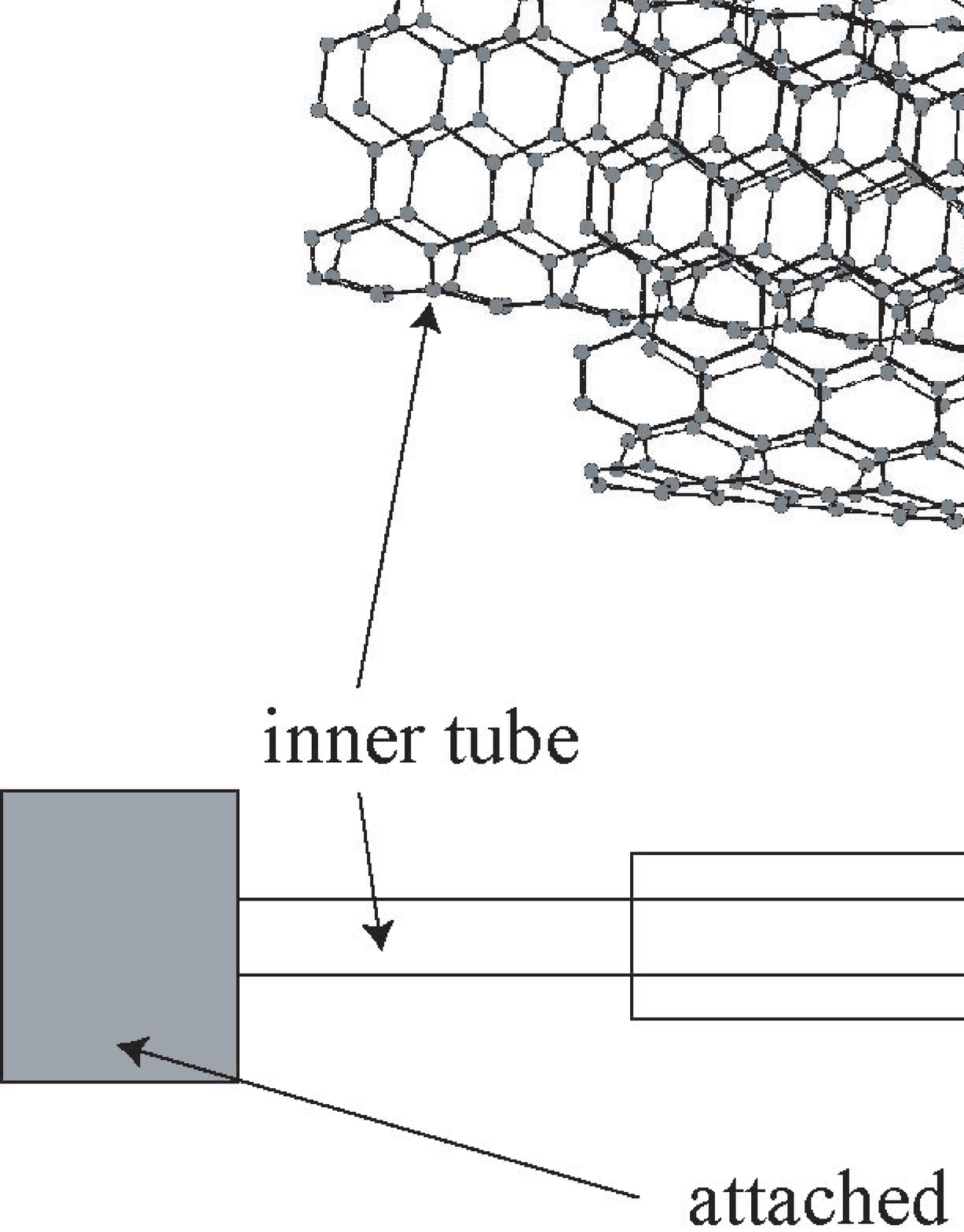} }}
\caption{
TDWNT }
\end{figure}

\begin{figure}
  \epsfxsize=0.7\columnwidth
\centerline{\hbox{   \epsffile{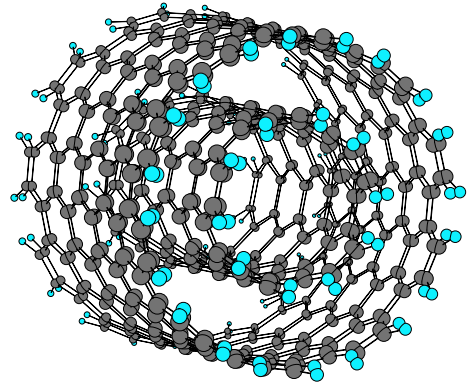} }}
\caption{
Structure [fi] superimposed on its initial  structure. }
\end{figure}

\begin{figure}
  \epsfxsize=0.5\columnwidth
\centerline{\hbox{   \epsffile{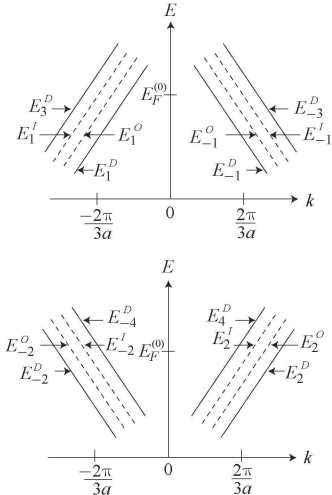} }}
\caption{
The dispersion lines near the neutral Fermi level $E_F^{(0)}$ 
 are numbered based on the index $\tau$ under the conditions $E_{|\tau|}^{\mu}\left( (-1)^\tau\frac{2\pi}{3a} \right)
 = E_{-|\tau|}^{\mu}\left(-(-1)^\tau\frac{2\pi}{3a} \right) \simeq E_F^{(0)}$
 and $ \tau \frac{dE_{\tau}^{\mu}}{dk}>0$.
 }
\end{figure}

\begin{figure}
  \epsfxsize=0.7\columnwidth
\centerline{\hbox{   \epsffile{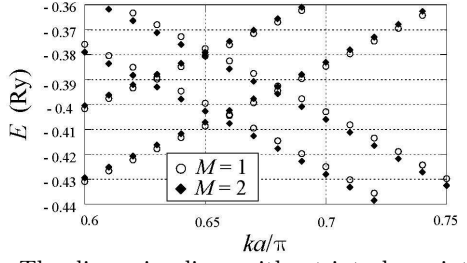} }}
\caption{
The dispersion lines 
 without   interlayer interaction   for $M=1$ (open circles)
  and $M=2$ (closed diamonds),
  where $M$ denotes the integer parameter  of Eq. (\ref{def-X(k)}).
}
\end{figure}

\begin{figure}
  \epsfxsize=0.7\columnwidth
\centerline{\hbox{   \epsffile{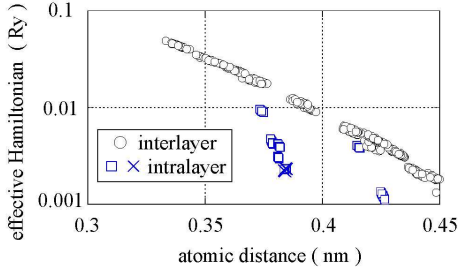} }}
\caption{Effective Hamiltonian elements defined by Eq. (\ref{def-effective-h}).}
\end{figure}

\begin{figure}
  \epsfxsize=0.7\columnwidth
\centerline{\hbox{   \epsffile{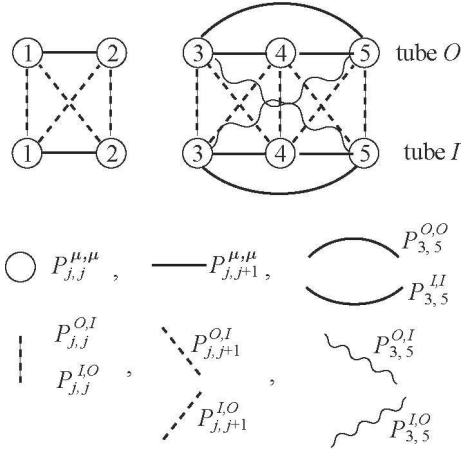} }}
\caption{
Schema for the  matrix $P$.
}
\end{figure}

\begin{figure}
  \epsfxsize=0.7\columnwidth
\centerline{\hbox{   \epsffile{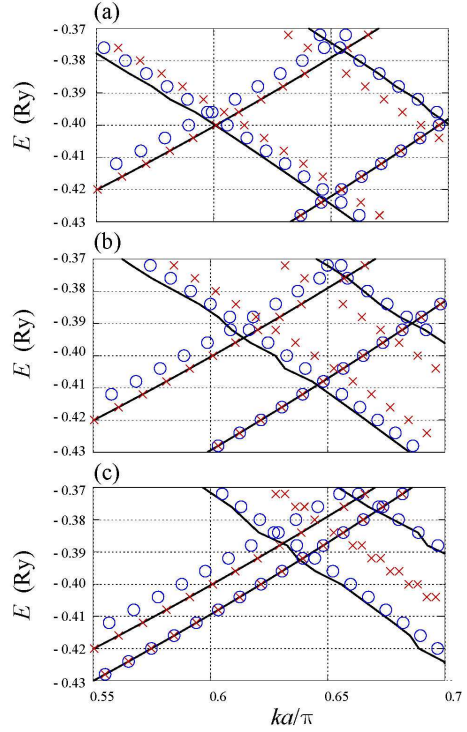} }}
\caption{
 The dispersion lines
  for   (a) $\Delta\varepsilon = -0.015$ (b) $\Delta\varepsilon =$ 0 and (c)  $\Delta\varepsilon =$ 0.020  Ry.
 At $k=0.65\pi/a$,
  $\varepsilon_{-1}= -0.022+\Delta\varepsilon$
 and
  $\varepsilon_{2}= -0.028+\Delta\varepsilon$.
 The crosses represent  $k^O_{\tau}(E)$ and $k^I_{\tau}(E)$ 
{\it without}  interlayer interaction.
The dispersion lines {\it with} interlayer interaction
 are shown by circles  for the exact $k^D(E)$ and  by solid lines
  for the approximate  $k^{D,A}(E)$.
 The wavenumber $k$ is calculated for the discrete energies
 $E=-0.428+0.004j$ where $j=0,1,2,\cdots, 14$.
}
\end{figure}

\begin{figure}
  \epsfxsize=0.7\columnwidth
\centerline{\hbox{   \epsffile{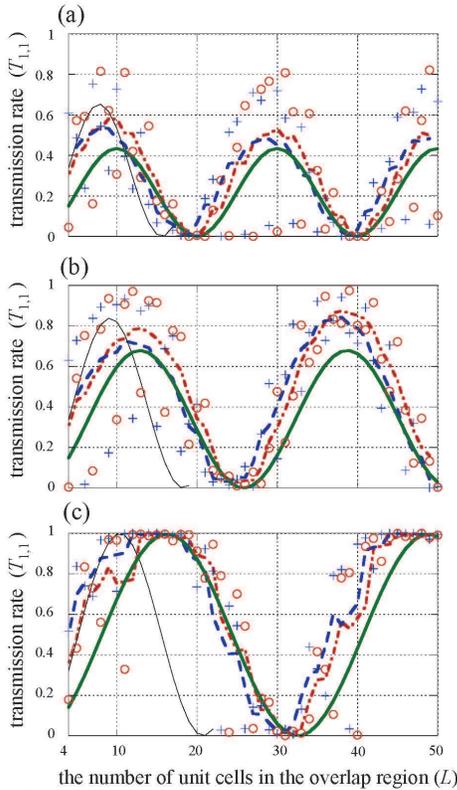} }}
\caption{
The transmission rate $T_{1,1}$  at an energy $E= -0.4$  Ry.
The values of $\Delta\varepsilon$ are the same as in Fig.   7.
The circles represent 
the exact $T_{1,1}$ with Eqs.  (\ref{IDD}) and (\ref{DDO}).
The crosses  represent the exact $T_{1,1}$
 when $q_0'$ and $q_0''$ in Eqs.  (\ref{IDD}) and (\ref{DDO})
 are replaced by $Q_0^D$.
The average defined by $T^{ave}(L)=(T(L-1)+T(L)+T(L+1))/3$ are 
 shown by the dot-dashed and dashed lines for the circles
 and crosses, respectively.
 The approximate formula (\ref{T'}) 
 is  shown by thick solid lines.
The thin solid represent Eq.(\ref{T'})
 where $E\widetilde{B}^{O,I}$ in  Eq.(\ref{def-A}) is replaced by zero,
 i.e.,  the interlayer overlap matrix is set to zero.
 The thin solid lines are shown only for the first period.
}
\end{figure}

\begin{figure}
  \epsfxsize=0.7\columnwidth
\centerline{\hbox{   \epsffile{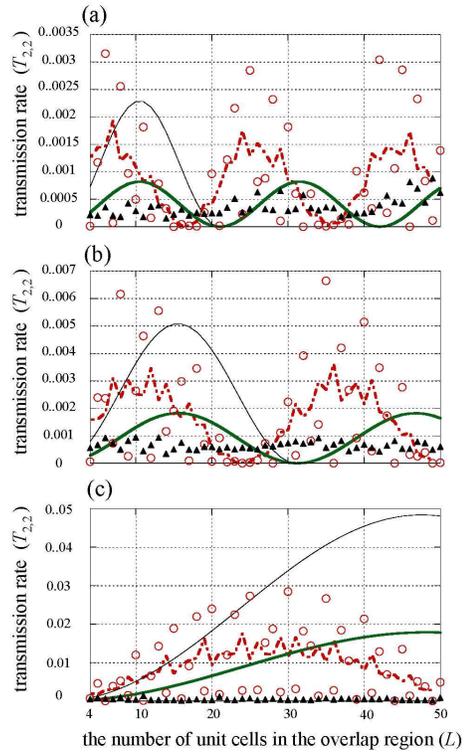} }}
\caption{
The same calculation as in Fig.8 for the transmission rate $T_{2,2}$.
 The crosses and dashed lines are omitted.
 The triangles show Eq. (\ref{err(S)}).
}
\end{figure}


\begin{thebibliography}{10}
\bibitem{note}
 In the input file for the generation
 of  the pseudopotential,  following items are used.
The type of  exchange-correlation is  Ceperley-Alder.
 The generation scheme is improved Troullier-Martins.
 The core radii for the  carbon  2s, 2p, 3d, 4f
orbitals  are 1.50, 1.54, 1.50 and  1.50 bohr, respectively.
 Those for the  hydrogen  1s, 2p, 3d, 4f orbitals
 are 1.25 bohr.
\end{thebibliography}
\end{document}